\documentclass[aps,prl,amsfonts,twocolumn,nofootinbib]{revtex4-1}
\usepackage{graphicx,latexsym}

\def\bea{\begin{eqnarray}}
\def\eea{\end{eqnarray}}
\def\be{\begin{equation}}
\def\ee{\end{equation}}
\newcommand{\ub}[1]{\underline{#1}}

\begin{document}

\title{BRST-invariant Pauli--Villars regularization of QCD}

\author{Sophia S. Chabysheva}
\author{John R. Hiller}
\affiliation{Department of Physics and Astronomy\\
University of Minnesota-Duluth \\
Duluth, Minnesota 55812}

\date{\today}

\begin{abstract}

We extend the QCD Lagrangian to include Pauli-Villars (PV) gluons,
quarks, and ghosts in such a way as to retain BRST invariance
in an arbitrary covariant gauge.
The extended Lagrangian can provide a starting point for
nonperturbative calculations in QCD, particularly with light-front
techniques, and the methods used
to construct it may be useful for perturbative calculations
in theories where dimensional regularization is not viable.
The regularization is arranged by having all interaction
terms in the Lagrangian be couplings between null fields,
specific combinations of positive and negative-metric PV fields.
The construction is done in steps, beginning with a gauge-invariant
Lagrangian with massless PV gluons and degenerate-mass PV quarks.
Auxiliary scalars are introduced to give mass to the PV gluons,
and break the mass degeneracy of the PV quarks, following a method
due to Stueckelberg.  Gauge fixing terms for the gluon 
fields are a part of this construction.  The ghost terms are
then obtained and shown to provide the BRST invariance.
A lack of dependence on the gauge parameter can be checked by
the calculation of physical quantities for a range of values of
the parameter.

\end{abstract}

%
\pacs{12.38.Lg, 11.15.Tk, 11.10.Ef, 11.10.Gh
}

\maketitle

The principal goal of hadronic physics is the nonperturbative solution
of quantum chromodynamics (QCD), in order to compute properties of
hadrons as bound states of quarks and gluons.  Lattice gauge theory~\cite{lattice}
is one approach for doing this; it has met with considerable success, but is limited
by its reliance on a Euclidean framework and by other difficulties, such as
fermion doubling and a large pion mass.  
Dyson--Schwinger methods~\cite{DSE} are also useful but also Euclidean.  The 
truncated conformal space approach~\cite{Hogervorst:2014rta} shows promise but is in its infancy.
An  alternative, which retains a Minkowski framework and the intuitive
notion of wave functions, is the light-front Hamiltonian approach~\cite{LFreview,Glazek}.
Light-front methods have been successfully applied to many gauge field theories,
including quantum electrodynamics~\cite{LFQED,Vary,ArbGauge}, QCD in two dimensions~\cite{2Dqcd},
and supersymmetric Yang--Mills theories~\cite{SDLCQ}; however, with respect to
four-dimensional QCD, this approach has always lacked a consistent regularization
that can be used nonperturbatively.

The standard regularization used for non-Abelian theories such as QCD is 
dimensional regularization~\cite{dimreg}.  Unfortunately, this method is inherently
perturbative, relying as it does on the analysis of the singularity structure of
Feynman diagrams and the modification of the associated integrations.  All of
this is buried deep within a nonperturbative calculation, where such separations
and modifications are difficult if not impossible to perform.

A regularization that can be introduced at the Lagrangian or Hamiltonian
level is much to be preferred, and there has been just such an approach
for Abelian theories, developed in the context of light-front Hamiltonian
methods~\cite{LFYukawa,Karmanov,LFQED}.\footnote{There has also been work on 
construction of a light-front regularization that is perturbatively 
equivalent to covariant regularization~\protect\cite{Pastonetal}.}
Massive Pauli--Villars (PV) fields are introduced 
to the Lagrangian with couplings arranged to provide the cancellations
necessary for regularization via subtractions of negative-metric
PV contributions, and the Hamiltonian derived in some fixed gauge.  

The cancellations are obtained by constructing all interactions from
combinations of positive and negative-metric fields that are null
in the following sense.  Let the (light-front) mode expansions for the gluon and
quark fields be written as
\bea
A_{ak\mu}(x) &=& \int \frac{d\ub{q}}{\sqrt{16\pi^3}}
          \sum_\lambda e_\mu^{(\lambda)}\left[a_{ak\lambda}(\ub{q})e^{-iq\cdot x} \right. \\
    && \rule{1.1in}{0mm}   \left.   +a_{ak\lambda}^\dagger(\ub{q})e^{iq\cdot x}\right], \nonumber \\
\psi_i(x)&=&\int \frac{d\ub{q}}{\sqrt{16\pi^3 q^+}}
       \sum_{as} \left[ u_{ais}(\ub{q})b_{ais}(\ub{q})e^{-iq\cdot x} \right. \\
    && \rule{1.1in}{0mm}   \left.    + v_{ais}(\ub{q}) d_{ais}^\dagger(\ub{q})e^{iq\cdot x}\right], \nonumber
\eea
with $d\ub{k}=dk^+ d^2k_\perp$ the light-front volume element~\cite{LFreview},
$e_\mu^{(\lambda)}$ a set of polarization vectors, $a_{ak\lambda}^\dagger$
the gluon creation operator with color index $a$, PV index $k$, and polarization $\lambda$, and
$b_{ais}^\dagger$ the quark creation operator with color index $a$, PV index $i$, and spin $s$.
The quark flavor index is suppressed, as it will be throughout.  
The physical fields are denoted by a PV index of zero.
The (anti)commutation relations for the creation operators are
\bea
{[}a_{ak\lambda}(\ub{q}),a_{bl\lambda'}^\dagger(\ub{q}')]
&=&r_k\epsilon^\lambda\delta_{ab}\delta_{kl}\delta_{\lambda\lambda'}\delta(\ub{q}-\ub{q}'), \\
{\{}b_{ais}(\ub{q}),b_{bjs'}^\dagger(\ub{q}')\}
&=&s_i \delta_{ab}\delta_{ij}\delta_{ss'}\delta(\ub{q}-\ub{q'}).
\eea
Here $r_k=\pm1$ and $s_i=\pm1$ are metric signatures for the gluon and quark fields with
PV index $k$ and $i$, respectively.  In addition, the gluon field has a polarization
dependent metric signature specified by $\epsilon^\lambda\equiv(-1,1,1,1)$.  Null
fields are then constructed as
\be
A_a^\mu\equiv \sum_k \xi_k A_{ak}^\mu,\;\; 
\psi\equiv \sum_i \beta_i \psi_i,
\ee
where the $\xi_k$ and $\beta_k$ are coupling coefficients constrained
to satisfy
\be \label{eq:xik}
\sum_k r_k \xi_k^2=0, \;\;
\sum_i s_i \beta_i^2=0.
\ee
Obviously, at least one PV gluon and one PV quark must have a negative
metric.  Typically one would take $r_k=(-1)^k$ and $s_i=(-1)^i$.

Interaction terms, such as $g\bar\psi\gamma^\mu T_a A_{a\mu}\psi$, are then
built from null combinations, with $r_0=s_0=\xi_0=\beta_0=1$
for the physical field. Here $T_a$ is a matrix representing the Lie
algebra, and repeated color indices are summed.  Any loop must then contain 
the sum $\sum_{ik} s_i \beta_i \beta'_i r_k \xi_k \xi'_k$, with the primed 
coefficients coming from the second vertex, which may or may not be 
the same as the first.  The metric signatures come from contraction
of the creation and annihilation operators associated with the internal
lines of the loop.  If the two vertices are the same, the null constraints immediately
provide two subtractions; if not the same, the different null combinations 
from the different vertices must be mutually null, that is $\sum_i s_i \beta_i \beta'_i =0$
and $\sum_k r_k \xi_k \xi'_k=0$, to again provide two subtractions.

This approach has been used to study QED in an 
arbitrary covariant gauge, including study of
the dependence of physical quantities on the gauge parameter~\cite{ArbGauge}.
The couplings induce currents that change the PV index of the field,
including mixing with the physical field, which breaks gauge invariance;
however, in the limit that the PV fields are removed, gauge invariance
is restored.

The extension to non-Abelian theories is, of course, problematic.  Unlike
Abelian theories, where a massive vector boson is known to be 
renormalizable~\cite{MassiveVector}, the proof of renormalizability
for Yang--Mills theories is best approached in terms of BRST invariance~\cite{BRST},
which would appear to require an underlying gauge invariance, which in turn
requires massless vector bosons.  The well-known exception is the use of spontaneous
symmetry breaking to provide mass to gauge bosons without destroying
gauge invariance at the Lagrangian level, which is a hint as
to how one might proceed.  Another difficulty is that, even for massless
PV gluons and mass-degenerate PV quarks, the null couplings that mix fields
do break gauge invariance.

We have resolved these difficulties.  The giving of mass to the PV gluons and
the lifting of the PV-quark masses is accomplished by a non-Abelian generalization
of Stueckelberg's mechanism~\cite{Stueckelberg,Marnelius:1997rx}, which associates
an auxiliary real scalar with each gluon field, simultaneously giving mass to the 
gluon and the scalar while also fixing the gauge.  The breaking of gauge invariance 
by the field-mixing interactions is eliminated by generalizing the definition of the 
gauge transformation to also include field mixing; the original gauge transformation 
is recovered in the limit that the PV fields are removed, by taking their masses
to infinity.  The BRST invariance is recovered for finite PV masses by inclusion 
of the appropriate Faddeev--Popov~\cite{FaddeevPopov} ghost terms required by the 
gauge-fixing terms.

In case this sounds too good to be true, we must point out that there are
drawbacks to our approach.  One is the presence of nonlocal interactions
for the ghost fields; this is made necessary by the mechanism used to
generate PV gluon masses.  The other drawback is a proliferation of
PV fields, which will be a large computational burden.  There are
several constraints that the PV couplings must satisfy, and each
constraint requires the addition of one or two PV fields.  The
formulation given here can require as many as four PV gluons, three PV quarks,
five PV adjoint real scalars, and four PV ghosts and anti-ghosts.  This
is for each color and flavor, and therefore is a very large number.

Oddly enough, the Higgs mechanism is not used.  We found that the requirement
of null couplings, to provide the regularization, causes two difficulties.
One is in the Higgs sector itself, where the null $\phi^4$ interaction
does not lead to a well-defined minimum to drive the breaking of the
symmetry.  The other is that gauge-invariant null couplings of the PV Higgs
to the gluons requires that any symmetry breaking leave as
massless only a null combination of gluon fields, rather than produce a
massless physical gluon and massive PV gluons.

In what follows, we describe the construction in stages.  The first is a Lagrangian 
for massless gluons and mass-degenerate quarks that is invariant with respect
to a generalized gauge transformation.  Next, we introduce the auxiliary scalars 
and add terms that give mass to the PV gluons and fix the gauge.  This
is followed by specification of a term that lifts the mass degeneracy
of the PV quarks.  Finally, we determine the Faddeev--Popov ghost term
and state the BRST transformations of the fields.
Our construction is kept slightly more general than QCD, in that we
treat $SU(N)$ Yang--Mills theory with fundamental matter, with QCD
being the $N=3$ case.  

The gauge-invariant Lagrangian for massless PV gluons and mass-degenerate 
PV quarks is
\bea \label{eq:Lbase}
{\cal L}&=&-\frac14\sum_k r_k F_{ak}^{\mu\nu} F_{ak\mu\nu}
  +\sum_i s_i \bar\psi_i(i\gamma^\mu\partial_\mu-m)\psi_i  \nonumber \\
  && +g\sum_{ijk}\beta_i \beta_j \xi_k \bar\psi_i\gamma^\mu T_a A_{ak\mu}\psi_j.
\eea
The quark spinor field $\psi_i$ is a column vector with respect to color. 
The quarks (of a single flavor) are all of mass $m$.
The gluon field tensor $F_{ak}^{\mu\nu}$ is computed from the field as
\be
F_{ak}^{\mu\nu}=\partial^\mu A_{ak}^\nu-\partial^\nu A_{ak}^\mu
      -r_k \xi_k gf_{abc}\sum_{lm}\xi_l\xi_m A_{bl}^\mu A_{cm}^\nu.
\ee
The structure constants $f_{abc}$ specify the commutation relation 
$[T_a,T_b]=if_{abc}T_c$.  

The gauge transformations of the fields are
\bea
A_{ak}^\mu &\longrightarrow & A_{ak}^\mu+\partial^\mu\Lambda_{ak}
+r_k\xi_k g f_{abc}\Lambda_b A_c^\mu, \\
\psi_i &\longrightarrow & \psi_i+ig s_i\beta_i T_a\Lambda_a \psi,
\eea
with $\Lambda_a\equiv\sum_k\xi_k\Lambda_{ak}$.
The null combination $\psi$ is then gauge invariant, and the
null combination $A_a^\mu$ is Abelian
\be
A_a^\mu \longrightarrow A_a^\mu+\partial^\mu\Lambda_a,
\ee
as is the associated field tensor
\be
F_a^{\mu\nu}=\sum_k \xi_k F_{ak}^{\mu\nu}=\partial^\mu A_a^\nu-\partial^\nu A_a^\mu.
\ee
The Lagrangian ${\cal L}$ is gauge-invariant with respect to these
transformations.

When expressed in terms of the null combinations, the Lagrangian reduces to
\bea \label{eq:Lbasenull}
{\cal L}&=&-\frac14\sum_k r_k(\partial^\mu A_{ak}^\nu-\partial^\nu A_{ak}^\mu)^2
                  +gf_{abc} \partial^\mu A_a^\nu A_{b\mu} A_{c\nu} \nonumber \\
          &&        +\sum_is_i\bar\psi_i(i\gamma^\mu\partial_\mu-m)\psi_i
                  +g\bar\psi\gamma^\mu T_a A_{a\mu}\psi.
\eea
All interactions are between null combinations, and the free part of the
Lagrangian corresponds to terms for massless gluon fields with metric signatures $r_k$
and degenerate-mass quark fields with signatures $s_i$.  The four-gluon interaction
that normally appears in the QCD Lagrangian has disappeared, as a result of cancellations
between physical and PV gluons.  It is restored between physical gluons in the
infinite-mass limit for PV gluons as a contraction of two three-gluon vertices,
with the contraction being a PV gluon.  This provides a mechanism, natural in the
context of the present Lagrangian, for the often-used trick of introducing an
auxiliary field to reduce the four-gluon interaction to two three-gluon interactions
for the convenience of color factors in perturbation theory~\cite{fourgluontrick}.

To give mass to the PV gluons, we use a non-Abelian extension of the Stueckelberg 
mechanism~\cite{Stueckelberg,Marnelius:1997rx}.  Real adjoint scalar fields 
$\phi_{ak}$ are introduced with a gauge transformation of 
\be
\phi_{ak} \longrightarrow \phi_{ak}+\mu_k \Lambda{ak}
    +\mu_k r_k \xi_k g f_{abc} \int^x \!\!\!\! dx'_\mu \Lambda_b(x') A_{c}^\mu(x').
\ee
The line integral is present in order that the gauge transformation of
the derivative be
\be
\partial^\mu\phi_{ak} \longrightarrow \partial^\mu\phi_{ak}+\mu_k \partial^\mu\Lambda{ak}
    +\mu_k r_k \xi_k g f_{abc} \Lambda_b A_{c}^\mu.
\ee
This then makes the combination $\mu_k A_{ak}^\mu-\partial^\mu\phi_{ak}$ gauge
invariant, which is used in the construction of the first term of an additional
piece for the Lagrangian, written as
\bea \label{eq:Lg}
{\cal L}_g&=&\frac12\sum_k r_k \left(\mu_k A_{ak}^\mu-\partial^\mu\phi_{ak}\right)^2 \\
  &&     -\frac{\zeta}{2}\sum_k r_k \left(\partial_\mu A_{ak}^\mu +\frac{\mu_k}{\zeta}\phi_{ak}\right)^2.
       \nonumber
\eea
The second term is the gauge fixing term.  However, when the two terms are
combined, the cross terms sum to a total divergence which can be neglected,
leaving
\bea \label{eq:Lgreduced}
{\cal L}_g&=&\frac12\sum_k r_k \mu_k^2 \left(A_{ak}^\mu\right)^2
           -\frac{\zeta}{2}\sum_k r_k \left(\partial_\mu A_{ak}^\mu\right)^2 \\
    &&       +\frac12\sum_k r_k\left[\left(\partial_\mu\phi_{ak}\right)^2
                                  -\frac{\mu_k^2}{\zeta}\phi_{ak}^2\right]. \nonumber
\eea
This is immediately seen to provide a mass $\mu_k$ and standard gauge-fixing term 
for each gluon field, with gauge parameter $\zeta$, and a free Lagrangian for
real scalars with masses $\mu_k/\sqrt{\zeta}$ and metric signatures $r_k$.  For the
physical gluon, the mass $\mu_0$ is to be taken to zero.  For pure Yang--Mills
theory, this can be done explicitly, by exclusion of the scalar fields $\phi_{a0}$
and replacement of the $k=0$ terms in ${\cal L}_g$ by a simple gauge-fixing
term for the physical gluons: $-\frac{\zeta}{2}\left(\partial_\mu A_{a0}^\mu\right)^2$.
When quarks are included, the $k=0$ scalar is needed, and the mass $\mu_0$
must be taken to zero as a limit; however, as will be seen, the scalar $\phi_{a0}$
does not couple to the physical fields for any value of $\mu_0$.

The mass degeneracy of the PV quarks can be removed by coupling the quarks
to a null combination of the scalars $\phi_{ak}$,
\be
\widetilde\phi_a\equiv\sum_k \xi_k\frac{\mu_{\rm PV}}{\mu_k}\phi_{ak},
\ee
made null by the additional constraint
\be \label{eq:xik/muk}
\sum_k r_k \frac{\xi_k^2}{\mu_k^2}=0.
\ee
Here $\mu_{\rm PV}\equiv \max_k \mu_k$ is the mass scale of the PV gluons.
The gauge transformation of the combination is Abelian
\be
\widetilde\phi_a \longrightarrow \widetilde\phi_a+\mu_{\rm PV}\Lambda_a
\ee
and provides the necessary piece to make gauge-invariant
the Lagrangian term
\be
{\cal L}_q=-\sum_i s_i m_i (\bar\psi_i+ig\frac{s_i\beta_i}{\mu_{\rm PV}}\widetilde\phi_a \bar\psi T_a)
                 (\psi_i-ig\frac{s_i\beta_i}{\mu_{\rm PV}}\widetilde\phi_a T_a\psi),
\ee
which is to replace the quark-mass term $-\sum_i s_i m \bar\psi_i\psi_i$
in (\ref{eq:Lbasenull}).  Thus, each PV quark can have a different mass $m_i$,
with $m_0$ the mass of the physical quark.

This new term can be written as
\bea \label{eq:Lq}
{\cal L}_q&=&-\sum_i s_i m_i \bar\psi_i\psi_i
            -ig \frac{m_{\rm PV}}{\mu_{\rm PV}}\left[\bar\psi T_a \widetilde\phi_a\widetilde\psi
                                             -\bar{\widetilde\psi} T_a \widetilde\phi_a\psi \right]
            \nonumber \\
    &&        -\sum_i s_i \frac{m_i \beta_i^2}{\mu_{\rm PV}^2}
                    \bar\psi T_a \widetilde\phi_a T_b\widetilde\phi_b\psi,
\eea
where we define a PV-quark mass scale $m_{\rm PV}\equiv\max_i m_i$
and a new null combination of quark fields
\be
\widetilde\psi=\sum_i \beta_i \frac{m_i}{m_{\rm PV}}\psi_i.
\ee
For this to be null, there is the additional constraint
\be
\sum_i s_i m_i^2 \beta_i^2=0,
\ee
and for $\widetilde\psi$ and $\psi$ to be mutually null, we must have
\be
\sum_i s_i m_i \beta_i^2=0.
\ee
This last constraint has the advantage of making the last term in
(\ref{eq:Lq}) simply zero.  Also, the second and third terms combined
are proportional to $\sum_{ij}\beta_i\beta_j (m_j-m_i)\bar\psi_iT_a\widetilde\phi_a\psi_j$,
which is zero for $i=j$ and guarantees that none of the scalars are directly
coupled to the physical quarks, as claimed above.  There are now three
constraints on the PV quark masses and coupling coefficients, which 
requires three PV quarks if all masses $m_i$ are to be chosen independently.

The last step of the construction is the Faddeev--Popov ghost term,
which can be done by the usual methods of path-integral quantization~\cite{BailinLove}.
We obtain, for ghosts $c_{ak}$ and anti-ghosts $\bar c_{ak}$,
\bea \label{eq:LFP}
\lefteqn{{\cal L}_{\rm FP}=\sum_k r_k \partial_\mu \bar c_{ak}\partial^\mu c_{ak}
                  -\sum_k r_k \frac{\mu_k^2}{\zeta} \bar c_{ak} c_{ak} }&&\\
       &&           +g f_{abc}\left[\partial_\mu \bar c_a c_b A_c^\mu
                  -\frac{\mu_{\rm PV}^2}{\zeta}\bar{\widetilde c}_a
                       \int^x\!\!\!\! dx'_\mu c_b(x') A_c^\mu(x')\right],
                  \nonumber
\eea
with the null combinations defined as
\be
c_a\equiv\sum_k\xi_k c_{ak},\;\;
\bar c_a\equiv\sum_k\xi_k \bar c_{ak}, \;\;
\bar{\widetilde c}_a\equiv\sum_k \xi_k\frac{\mu_k^2}{\mu_{\rm PV}^2}\bar c_{ak}.
\ee
For these to be (mutually) null, we require
\be
\sum_k r_k \mu_k^2 \xi_k^2=0,\;\;
\sum_k r_k \mu_k^4 \xi_k^2=0.
\ee

There are now four constraints on the PV-gluon masses $\mu_k$ and coupling coefficients $\xi_k$:
the two above plus (\ref{eq:xik}) and (\ref{eq:xik/muk}).  If all the PV-gluon masses are to
be chosen independently, this would require four PV gluons; for constrained mass values, the number
of PV gluons can be less.  The number of PV ghosts is the same as the number of PV gluons;
the number of PV scalars is one more, except for pure Yang--Mills theory, where $\phi_{a0}$
is not needed.  Because $\mu_0\ll\mu_{\rm PV}$, there could be difficulty
with significant figures in numerical solutions of these constraints.  However, calculations
in QED~\cite{ArbGauge} have indicated that the $\mu_0/\mu_{\rm PV}$ ratio need not
be extremely small, so that this potential difficulty may be mitigated.

This completes the construction.  The full expression for the PV-regulated QCD Lagrangian 
can be obtained by adding ${\cal L}_{g}$ and ${\cal L}_{\rm FP}$,
found in (\ref{eq:Lgreduced}) and (\ref{eq:LFP}), to ${\cal L}$ in (\ref{eq:Lbasenull}),
and replacing the quark mass term in ${\cal L}$ with ${\cal L}_q$, found
in (\ref{eq:Lq}).  

The associated BRST transformations are determined from the 
gauge transformations with the replacement $\Lambda_{ak}\rightarrow\epsilon c_{ak}$,
where $\epsilon$ is a real Grassmann constant for which $\epsilon^2=0$.  The
transformations are
\bea
\delta A_{ak}^\mu&=&\epsilon\partial^\mu c_{ak}
                      +\epsilon r_k\xi_k g f_{abc}c_b A_c^\mu, \\
\delta \psi_i &=& i\epsilon g s_i \beta_i T_a c_a \psi, \;\;
\delta \bar\psi_i = -i\epsilon g s_i \beta_i \bar\psi T_a c_a, \\
\delta \phi_{ak}&=&\epsilon\mu_k c_{ak} \\
       &&               +\epsilon r_k\xi_k \mu_k g f_{abc}
                             \int^x\!\!\!\! dx'_\mu c_b(x') A_c^\mu(x'),
       \nonumber \\
\delta \partial^\mu\phi_{ak}&=&\epsilon\mu_k \partial^\mu c_{ak}
                      +\epsilon r_k\xi_k \mu_k g f_{abc} c_b A_c^\mu, \\
\delta \bar c_{ak} & =& -\zeta\epsilon\left(\partial_\mu A_{ak}^\mu+\frac{\mu_k}{\zeta}\phi_{ak}\right), \\
\delta c_{ak} &=& \frac12 \epsilon r_k \xi_k g f_{abc}c_b c_c.
\eea
For the various null combinations we then find
\bea
\delta A_a^\mu &=& \epsilon\partial^\mu c_a, \;\;
\delta \widetilde\phi_a= \epsilon\mu_{\rm PV}c_a, \;\;
\delta c_a = 0, \\
\delta \psi &=& 0, \;\;
\delta \widetilde\psi = 0.
\eea
All of the gauge-invariant pieces of the Lagrangian are then automatically
BRST invariant, and we need only check the sum of the second term of ${\cal L}_g$
and ${\cal L}_{\rm FP}$, in (\ref{eq:Lg}) and (\ref{eq:LFP}) respectively.
A short calculation shows that this sum, and therefore the entire Lagrangian,
is also BRST invariant.

However, what we have done is only a formal construction.  A next step is to construct 
the corresponding light-front Fock-space Hamiltonian and then to solve for its
eigenstates.  This will be facilitated by another aspect of null couplings,
in that the projections~\cite{LFreview} $\psi_-\equiv \frac12\gamma^0\gamma^-\psi$ and 
$\widetilde\psi_-\equiv\frac12\gamma^0\gamma^-\widetilde\psi$ of the Dirac field are gauge
invariant and satisfy constraint equations, projected from the Dirac equation,
that are independent of the gluon fields.  This allows for explicit solution
of the constraint without invocation of light-cone gauge~\cite{LFQED}.
There will also be cancellation of instantaneous fermion interactions,
just as seen in Yukawa theory~\cite{LFYukawa} and QED~\cite{LFQED}.
This should be a promising avenue of attack on QCD.

At the stage of solving the Fock-space eigenvalue problems, there is
a potential source of gauge dependence, in that most methods employ
Fock-space truncation to produce a finite computational problem.
A method that avoids such a truncation, the light-front coupled-cluster
method~\cite{LFCClett}, has been developed and should be applicable
to QCD.  Rather than truncate Fock space, a finite computational
problem is obtained by truncation of the way in which higher Fock-state
wave functions are related to the lower ones.



\end{document}